\documentclass[pra,showpacs,twoside,twocolumn,amsmath,amssymb]{revtex4}

\usepackage{graphicx}
\usepackage{mathtools} 
\usepackage[TABBOTCAP]{subfigure} 

\newcommand{\ket}[1]{\ensuremath{\vert{#1}\rangle}}                                     
                                     
\newcommand{\op}[1]{\ensuremath{\mathrm{#1}}}                                             
\newcommand{\hop}[1]{\ensuremath{\mathrm{\hat #1}}}                                             
\newcommand{\spa}[1]{\ensuremath{\mathcal{#1}}}                                         
\newcommand{\cha}[1]{\spa{#1}}                                                          
\newcommand{\dif}[2]{\ensuremath{\frac{\mathrm{d}}{\mathrm{d}#2}#1}}                    
\newcommand{\trace}{\mathrm{Tr}}                                                        
\newcommand{\pop}[1]{\ensuremath{\mathrm{\bar #1}}}                                     
\newcommand{\ii}{\ensuremath{\mathrm{i}}}                                            

\begin{document}

\title{Robustness of channel-adapted quantum error correction}

\author{G\'abor Ball\'o}
\email[E-mail: ]{ballo.g@freemail.hu}
\author{P\'eter Gurin}
\email[E-mail: ]{gurin@almos.vein.hu}
\affiliation{Institute of Physics, University of Pannonia,\\ H-8200 Veszpr\'em, Hungary}

\date{\today}

\begin{abstract}
A quantum channel models the interaction between the system we are interested in and its environment. Such a model can capture the main features of the interaction but because of the complexity of the environment we can not assume that it is fully accurate. We study the robustness of quantum error correction operations against completely unexpected and subsequently undetermined type of channel uncertainties. We find that a channel-adapted optimal error correction operation does not only give the best possible channel fidelity but it is more robust against channel alterations than any other error correction operation. Our results are valid for Pauli channels and stabilizer codes but based on some numerical results we believe that very similar conclusions can be drawn also in the general case.
\end{abstract}

\maketitle

\section{Introduction}
One of the central problems of the quantum information processing systems is to find suitable error correcting procedures. Therefore Shor's early result \cite{shor_9bit}, which was followed by the development of the standard theory of quantum error correction (QEC for short) \cite{knill,steane,gottesman}, was a very important step, giving the possibility for quantum information theory to become a potentially applicable science from an only theoretically interesting field. These standard quantum error correcting methods tried to adapt the techniques of classical information theory for the case of quantum information. However, it was later recognized that looking for the best error correcting method can be formulated as an optimization problem \cite{audenaert,reimpell,yamamoto}, and recently some authors approached the task of error correction from this new viewpoint \cite{fletcher,fletcher_near.opt,fletcher_phd,kosut}, and found that better solutions can be achieved than by the standard QEC. But there are relatively few studies about the comparison of these approaches in the aspect of resistance against the uncertainty of the channel model. In this paper we take a step in this direction. 

In QEC the Hilbert space of the system of interest is embedded into a Hilbert space of a larger system. This is the encoding. In the most simple case this means that a logical qubit is encoded into the strongly entangled states of a few physical qubits. After encoding the system goes through a noisy quantum channel then we apply a QEC operation to eliminate the effect of noise. The method of standard QEC is based on the assumption, that if the noise is weak, then those errors will be the most probable, which only corrupt the state of very few, perhaps only one qubit, therefore the standard error correction procedure is designed to correct perfectly these types of errors. This assumption is true under very general circumstances, and it is independent from the actual character of the channel. We only need to necessitate that the noise is acting on the qubits independently, but nothing else about the specific form of the channel. 

However, the channel models the interactions of the physical system with its environment, about which we can make plausible assumptions based on knowledge about the specific system. If we have information on the specific effect of the channel, then based on this we can design noise-specific encoding and error correction. The statement of error correction in this context, as an optimization problem only gained significant attention in the last few years \cite{fletcher,fletcher_near.opt,fletcher_phd,kosut}. They showed for example that by having knowledge about the exact effect of the channel, the error can be corrected with better efficiency, than by conventional QEC. A smaller system is enough for an equally effective encoding, and/or the code offers better protection. 

However it is also undoubted that in reality, because of the complexity of the environment, the interactions between the system and its environment can not be taken into account with perfect accuracy. The quantum channel will always remain just a model, which though describes reality more or less precisely, but we can not assume that it is fully accurate. So the question rises: if the noise differs in some degree from the chosen model, then how effective does our model-specific error correction method remain, i.e. how robust it is? 

In this work, our aim is the study of this question, that how robust are the conventional and optimization based channel specific error correction procedures against the uncertainties of the channel. At first sight we could think that the conventional QEC---which makes very slight assumptions about the properties of the channel---is more insensitive to the uncertainties, than an error correction optimized for a specific channel. On the other hand, we can also think that if a solution is optimal, then in general in case of the very small alteration of the conditions it decays only in second order, so it is robust in some measure. These questions apparently have greater significance when the channel is general, but because of the simplicity of the handling of Pauli channels, here we concentrate mainly on this specific case. 

The paper is organized as follows. In section \ref{sec:robustness.definition} we define the measure of robustness. In section \ref{sec:optimal.correction.for.pauli.channels} we demonstrate the procedure of searching the optimal correction operation for Pauli channels. In section \ref{sec:mixing.pauli.channels} we present the general model of robustness for Pauli channels. Finally, in section \ref{sec:conclusion} we make conclusions related to Pauli channels and based on numerical results we discuss the possible generalizations of these results to non-Pauli channels.

\section{Robustness}
\label{sec:robustness.definition}
A quantum channel is defined to be a completely positive and trace preserving (CPTP) map \cite{kraus}. This map can be represented by a set of operator elements, $\{\op{E}_i\}$, called the Kraus representation, which gives the channel output as
\begin{equation}
\cha{E}(\rho)=\sum_i\op{E}_i\rho\op{E}_i^\dagger\ ,
\label{eq:kraus.form}
\end{equation}
and the operator elements must satisfy the completeness relation 
\begin{equation}
\sum_i\op{E}_i^\dagger\op{E}_i=\hop{1}\ ,
\label{eq:tp.constraint}
\end{equation}
in order to represent a trace preserving map. We have to mention that the set of operator elements is not unique. 

First of all we have to characterize channel performance. The role of the entangled states in storing and transferring quantum information was explored long ago, and it is well established that the entanglement fidelity properly measures how successfully the channel preserves the entanglement, i.e. the quantum information \cite{schumacher}. But the entanglement fidelity depends not only on the channel $\cha{E}$, but also on the input state $\rho$. To avoid this, in this paper we always use the maximally entangled state as an input. The argument behind this is that if the channel can preserve the maximally entangled state, then it can preserve other entangled states also. This special case of the entanglement fidelity is sometimes called channel fidelity \cite{reimpell}, as it depends only on the properties of the channel. 
This can be given in terms of $\{\op{E}_i\}$ Kraus operators as
\[
 F_\mathrm{ch}(\cha{E})=\sum_i\left|\trace(\rho\op{E}_i)\right|^2=\sum_i\left|\frac{1}{\mathrm{dim}(\spa{H})}\trace(\op{E}_i)\right|^2\ ,
\]
where $\spa{H}$ is the Hilbert space of the system.

How effectively does a QEC operation $\cha{R}$ correct the errors which occur while the system goes through channel $\cha{E}$? As usual, in this paper we measure it with the performance of the composite channel $\cha{R}\circ\cha{E}$, that is with the fidelity $F_\mathrm{ch}(\cha{R}\circ\cha{E})$.
The aim of the optimization based channel specific error correction is thus to determine the error correction channel operation $\cha{R}^*$, for which $F_\mathrm{ch}(\cha{R}^*\circ\cha{E})$ is maximal, so the result certainly depends on the channel. The standard QEC also makes assumption about the channel model, but much less, as we have already mentioned in the introduction.

By fixing an encoding and a channel, we can compare the two methods. As a first step, we can compare the efficiency of the different error corrections, measuring them in terms of channel fidelity, certainly getting the optimization method always as the best. This however may not be striking, as if we know the nature of the error source, we can act against it more effectively.  

As a second step, we can compare the robustness of error correction operations. We can approach robustness from different aspects depending on how we model channel uncertainty. One possibility is to take the uncertainty into account in the formulation of the optimization problem, i.e. define the channel as an integral weighted by a probability distribution, then maximize the average entanglement fidelity. This method is shortly discussed by Fletcher in \cite{fletcher_phd}. A similar problem statement can be found in the work \cite{kosut}, which tries to solve the problem with a reformulated, ``indirect'' version of the optimization problem. Of course, to use these approaches we must first consider by what assumptions the distribution will be physically realistic. 
 
The drawback of such approaches is the exactly defined, ``well known'' manner of uncertainty. It can be very useful in special cases, but here we want to study the question of robustness from another side. How effectively can we protect information from noise, against which we are unprepared? We can seek answers for this question in the following way. We have a channel and a correction operation, then we alter the channel a bit, and check how effective the unaltered correction operation remains on the new channel. In this way we can get information on the robustness of a correction operation against completely unexpected and subsequently undetermined type of uncertainties. 

More formally, let $\cha{R}_0$ be an error correction operation originally used to correct the channel $\cha{E}_0$. Our aim is to characterize the robustness of $\cha{R}_0$ against the alteration of the channel. To alter the original channel, starting from $\cha{E}_0$, we move in the space of channels in a direction given by another channel $\bar{\cha{E}}$, in other words we \emph{mix} the first channel with the second one. Mathematically, we define channel mixing as the convex combination of channels. The operator elements of the mixed channel $\cha{E}'_\gamma$ are the following: 
\begin{equation}
\label{eq:mixed.channel.op.elements}
\{\op{E}'_i\}=\big\{\sqrt{1-\gamma}\op{E}_j\big\}\uplus\big\{\sqrt{\gamma}\pop{E}_k\big\}\ ,
\end{equation}
where $\{\op{E}_j\}$ and $\{\pop{E}_k\}$ are the sets of the operator elements of $\cha{E}_0$ and $\bar{\cha{E}}$, furthermore $\gamma\in[0,1]$ is the \emph{mixing parameter}. Notation $\uplus$ means here that if $\{\op{E}_j\}$ and $\{\pop{E}_k\}$ contains an element which is the same in these two sets then this common element will appear two times in $\{\op{E}'_i\}$. It is easy to see, that the obtained operator element set still represents a quantum channel, as channels form a convex set, so the trace preserving constraint \eqref{eq:tp.constraint} remains valid.

During channel mixing we observe the change in the effectiveness of $\cha{R}_0$ which is measured by $F_\mathrm{ch}(\cha{R}_0\circ\cha{E}'_\gamma)$. However, it is worth comparing it with the best possibility allowed by the given encoding and noisy channel $\cha{E}'_\gamma$, instead of just working with the pure value of it. Analyzing whether the effectiveness of $\cha{R}_0$ remains close to the best possible value or not while we alter the original channel by mixing it with some $\bar{\cha{E}}$, it turns out against which type of noise is the correction operation sensitive, and against which type it is robust. 

Thus the robustness of the $\cha{R}_0$ correction operation for a given $\cha{E}_0$ channel and $\bar{\cha{E}}$ alteration can be characterized with the difference between two entanglement fidelity values. First, we must determine the entanglement fidelity for the case when the effect of the mixed channel $\cha{E}'_\gamma$ is corrected by the best possible $\cha{R}^*_\gamma$ correction operation---which gives the biggest possible entanglement fidelity---, then we must check, compared to this optimal value, how much smaller is the entanglement fidelity we get by using the original $\cha{R}_0$ correction operation:
\begin{equation}
\label{eq:robustness.definition}
\Delta F_\mathrm{ch}(\gamma)=F_\mathrm{ch}(\cha{R}^*_\gamma\circ\cha{E}'_{\gamma})-F_\mathrm{ch}(\cha{R}_0\circ\cha{E}'_{\gamma})\ .
\end{equation}
This difference can be greater than zero because of two reasons. Firstly, it is possible that the examined $\cha{R}_0$ correction operation is not even optimal for the $\cha{E}_0$ channel. The $F_\mathrm{ch}(\cha{R}^*_0\circ\cha{E}_0)-F_\mathrm{ch}(\cha{R}_0\circ\cha{E}_0)$ difference, which is independent from the mixing parameter $\gamma$, characterizes optimality, i.e. how effective the $\cha{R}_0$ correction for the $\cha{E}_0$ channel is. Secondly, it can be greater than zero, if $\cha{R}_0$ is not robust. We want to use the expression ``robustness'' to characterize the feature of the error correction operation, which shows how the efficiency of $\cha{R}_0$ changes with the mixing parameter. Thus it turns out that the robustness against the alteration of the channel can be measured by the inspection of the $\gamma$ dependence of the $\Delta F_\mathrm{ch}$ function. If $\Delta F_\mathrm{ch}$ grows fast with $\gamma$, then the $\cha{R}_0$ correction is sensitive to the $\bar{\mathcal{E}}$ alteration, or else when $\cha{R}_0$ remains closely as effective as it was for the $\cha{E}_0$ channel, it is robust. The robustness could be defined as the derivative $\dif{\Delta F_\mathrm{ch}}{\gamma}$, however it is more useful to characterize the robustness with a domain in which the efficiency of $\cha{R}_0$ is not worse than it is for the original channel plus a given value. The boundary of the $\delta$-robustness domain can be defined as the channel $\cha{E}'_{\gamma_\delta}$ where $\gamma_\delta$ is the maximal value, for which it is true that $\Delta F_\mathrm{ch}(\gamma)-[F_\mathrm{ch}(\cha{R}^*_0\circ\cha{E}_0)-F_\mathrm{ch}(\cha{R}_0\circ\cha{E}_0)]\leq\delta$, if $\gamma\leq\gamma_\delta$. The whole $\delta$-robustness domain of a correction operation $\cha{R}_0$ around the channel $\cha{E}_0$ consists of all the channels which are inside the boundary.

To determine the boundary of the robustness domain we need to know the second term of \eqref{eq:robustness.definition}, i.e. how good entanglement fidelity we get for the channel $\cha{E}'_{\gamma}$ using the $\cha{R}_0$ correction operation. It can be seen, that we do not need optimization to determine this term. Using the operator elements of the mixed channel $\cha{E}'_{\gamma}$ as in \eqref{eq:mixed.channel.op.elements}: 
\begin{equation}
\label{eq:fidelity.linear.term}
     F_\mathrm{ch}(\cha{R}_0\circ\cha{E}'_{\gamma})
  =  (1-\gamma) F_\mathrm{ch}(\cha{R}_0\circ\cha{E}_0)
     +\gamma F_\mathrm{ch}(\cha{R}_0\circ\bar{\cha{E}})\ .
\end{equation}
That is, this term varies linearly in function of $\gamma$ as the distance from the channel $\cha{E}_0$ grows. This follows from our definition of channel mixing and the linearity of the trace.

More difficult it is to handle the first term of eq. \eqref{eq:robustness.definition}. For a given channel, the correction operation which gives the greatest entanglement fidelity in general can only be determined by numerical methods, which should most obviously produced as the solution of a semidefinite programming problem. In the literature authors mostly follow this approach \cite{yamamoto,audenaert,fletcher}, although there are also alternate attempts \cite{fletcher_near.opt,kosut}. Eventually, in the case of a special class of channels---the Pauli channels---the optimal correction operation can be generated analytically \cite{fletcher_phd,fletcher_near.opt}. This is a very essential result, as many important channels belong to this class, furthermore it also can help understand more complex cases, where only numerical methods are available.

\section{Optimal QEC for Pauli channels}
\label{sec:optimal.correction.for.pauli.channels}

\subsection{The standard QEC procedure}
The Pauli group on $n$ qubits is denoted by $G_n$. It is generated by all $3n$ one qubit Pauli operators $X_i$, $Y_i$, $Z_i$ acting on the $i$th qubit. It has order $2^{2n+2}$, because for each $i$ we can have $I$, $X_i$, $Y_i$ or $Z_i$, plus a possible overall factor of $\pm 1$ or $\pm \ii$.
If a channel has a set of operator elements which contains scaled elements of the Pauli group, i.e. $\{\op{E}_i=\sqrt{a_i}g_i\}$ where $a_i \geq 0$ and $g_i\in G_n$ , then we call it Pauli channel. Because of the trace preserving condition of the channel, $\sum_i a_i=1$ has to hold.

The QEC procedure begins with an encoding. This can be seen as a mapping $\op{U}_\spa{C} : \spa{H} \rightarrow \spa{K}$ where $\spa{H}$ is the Hilbert space of the system which is the source of the quantum information---it is spanned by the original qubits---, $\spa{K}$ is the Hilbert space of a larger system, and $\op{U}_\spa{C}$ is the encoding isometry which defines the code space $\spa{C} = \op{U}_\spa{C}(\spa{H}) \subset \spa{K}$ spanned by the logical qubits. The Knill--Laflamme theorem \cite{knill} states that the channel $\cha{E}$ is correctable \emph{perfectly} on $\spa{C}$ iff there exists a set of operator elements $\{\op{E}_j\}$ of $\cha{E}$  which satisfies the equation $\op{P}_\spa{C} \op{E}_i^\dagger \op{E}_j \op{P}_\spa{C} = c_{i} \delta_{ij} \op{P}_\spa{C}$, where $\op{P}_\spa{C}$ is the orthogonal projection onto $\spa{C}$ and the $c_{i}$ are scalars. It means that all the syndrome subspaces $\spa{S}_i \coloneqq \op{E}_i(\spa{C})$ are mutually orthogonal and the effect of $\op{E}_i$ after the encoder can be written as $\op{E}_i \op{U}_\spa{C}=\sqrt{c_{i}} \op{A}_i \op{W}_i \op{U}_\spa{C}$ where $\op{W}_i$ is an isometry between the code and the syndrome subspaces and $\op{A}_i$ is a unitary operator. In this case the errors can be corrected by a projective syndrome measurement followed by a unitary operation which depends on the result of the syndrome measurement. Actually this is $(\op{A}_i \op{W}_i\op{U}_\spa{C})^\dagger$, the inverse of the encoding operation followed by the error operator.

The efficiency of the  QEC depends on the code construction. Certainly the error operators of a general channel do not meet the requirements of the Knill--Laflamme conditions. The task is to find an encoding operation for which the most probable errors of a given channel prove to be correctable. But if the noise affects each qubit independently, i.e. the interaction operator between the qubits and their environment is tensor product of single-qubit operators, furthermore the noise is weak, then the most probable errors are those which affect only one qubit, therefore the aim is to find codes on which these errors can be corrected. We call this strategy in this paper the standard strategy of QEC. The success of the standard strategy does not depend on the actual form of the noise. It was the method of the early results of QEC, like e.g. the five qubit code \cite{laflamme,divincenzo}, and up to now this is the main stream of the QEC research.

\subsection{The optimal QEC construction}
As it can be seen from the previous subsection, after fixing the errors required to be corrected and constructing the code, the correction operator is determined. This is quite clear in case of stabilizer codes. The generators of the stabilizer group determine not only the code subspace, but also the syndrome subspaces as the intersection of the $\pm 1$ eigenspaces of each generator and the $\op{A}_i$ unitaries are determined by the implicit assumption that the most probable errors required to be corrected are the one qubit errors.

The case of Pauli errors and stabilizer codes is very special because every Pauli group error rotates the code subspace onto one of the mutually orthogonal syndrome subspaces \cite{gottesman_phd}. This is because all the elements of the Pauli group are either commute or anticommute with each other. In the case of an $[[n,k]]$ stabilizer code this means that the code subspace $\spa{C}$ which is stabilized by the subgroup $S=\langle g_1,\dots,g_{n-k}\rangle < G_n$ is rotated by the operator $g \in G_n$ onto the subspace stabilized by $\langle (-1)^{q_1}g_1,\dots,(-1)^{q_{n-k}}g_{n-k}\rangle$, where $q_i = 0$ if $[g,g_i] = 0$ and $q_i = 1$ if $\{g,g_i\} = 0$. It is convenience to label these syndrome subspaces as $\spa{S}_q$ by the indices $q=q_{n-k}2^{n-k-1}+\dots+q_12^0$. 

It is clear that the class of Pauli operators which map $\spa{C}$ onto itself form the $N(S)$ normalizer of $S$ which is also a subgroup in $G_n$. Furthermore, all the Pauli operators from a $g \cdot N(S)$ left coset maps $\spa{C}$ onto the same syndrome subspace $\spa{S}_q$. Let us fix a representative element from each of these cosets and denote them by $\op{W}_q$. After this, an arbitrary element $g$ of the Pauli group can be written as $g = n \op{W}_q$ where $n \in N(S)$ and $\op{W}_q$ are uniquely determined. The stabilizer is a subgroup in its own normalizer, $S < N(S)$. Therefore by fixing a representative element $\ii^r \op{A}_p$ from each of the left cosets of $S$ which are contained by the normalizer, i.e. $\ii^r \op{A}_p S \subset N(S)$, we can write $n = \ii^r \op{A}_p s$, where $r$, $p$ and $s \in S$ are uniquely determined for each $n$. Here we write down the scalar factor $\ii^r$ explicitly, ($r=0,...3)$, which can be done because if $\op{A}_p S \subset N(S)$ then $\ii^r \op{A}_p S \subset N(S)$ too. Actually, the $\op{A}_p$ normalizer operators are the logical Pauli operators which act on the logical qubits as the Pauli operators on the physical qubits therefore they are self-adjoint operators. We conclude that an arbitrary Pauli error operator can be written as
\begin{equation}
\label{eq:pauli.op.form}
     \op{E}= \sqrt{a}\, g = \sqrt{a}\; \ii^r \op{A}_p \op{W}_q s
\end{equation}
where $s \in S$, $p=0, \dots , 2^{2k}-1$ and $q=0, \dots , 2^{n-k}-1$ are uniquely determined by $\op{E}$. The index $r$ is unimportant, without loss of generality we can assume that it is zero because the $\ii^r \op{E}$ operators transform the density operator the same as $\op{E}$, as can be seen from eq. \eqref{eq:kraus.form}. So, for a given encoder, we classified the Pauli error operators. All these are in one of the left cosets of the stabilizer, $\op{A}_p \op{W}_q S$, where $q$ means that the error maps $\spa{C}$ onto $\spa{S}_q$ and $p$ specifies a further unitary operation.

After this classification of the error operators let us see the error correction. All of the operators from the same coset $\op{A}_p \op{W}_q S$ transform the code subspace equivalently, that is they cause the same error which can be corrected by the same error correcting operation. (When there are more than one important error operators in the same coset which are required to be corrected, then it is the case of degenerate QEC.) Therefore it is convenient to reduce the error operators of the form \eqref{eq:pauli.op.form} with a given $p$ and $q$ into one operator $\sqrt{a_{p,q}} \op{A}_p \op{W}_q$, where $a_{p,q} = \sum_i a_i$, and the summation index runs over all $i$ for which $\op{E}_i$ can be written as \eqref{eq:pauli.op.form} with the given $p$ and $q$.

Due to the effect of the error operator $a_{p,q} \op{A}_p \op{W}_q$ the code moves to the syndrome subspace $\spa{S}_q$ and there it gets transformed further by the $\op{A}_p$ operator. Furthermore, this error occurs with probability $a_{p,q}$. Notice that if this is the only error operator in the channel (i.e. if $a_{p,q} =1$), then this channel is perfectly correctable with the $\op{U}_\spa{C}^\dagger \op{W}_q^\dagger \op{A}_p$ operator, as this is exactly the inverse of the encoding operation followed by the error operator. If there is another error operator in the channel, which is indexed by a different $q$ integer, then this means that with some probability, the code can get into some other syndrome subspace, which is orthogonal to the previous. This is not a problem, as---according to the spirit of the Knill--Laflamme theorem---after the syndrome measurement, the code is already projected into one of the $\spa{S}_q$ subspaces, so we can apply such a correction operation, which has two different operator elements, one corrects the first error, and the other corrects the second. But if more error operators of the channel move the code into the same syndrome subspace in different ways, then---as stated by the Knill--Laflamme theorem---all of these can not be corrected perfectly at the same time. It can be seen from the following that we can choose any of them, and---in case of Pauli errors---can also correct it. But only one $p$ for each $q$. Here it concludes, how effective the correction will be for a given channel: the correction will be the best, if for each $q$ we choose and correct the most probable error. In contrast, the standard QEC chooses to correct the few-qubit (more often only the single-qubit) errors.

Now we show, that the intuitive strategy above really gives the greatest entanglement fidelity. Let the index of the most probable error operator belonging to each $\spa{S}_q$ syndrome subspaces be denoted by $p_q^*$. Now the operator element set of the optimal $\cha{R}^*$ correction operation can be obtained in $\{\op{U}_\spa{C}^\dagger \op{W}_q^\dagger \op{A}_{p_q^*}\}_{q=0}^{2^{n-k}-1}$ form. Using this correction operation the entanglement fidelity: 
\begin{equation}
\label{eq:entanglement.fidelity.analytical.form}
\begin{split}
     F_\mathrm{ch}&(\mathcal{R}^*\!\circ\mathcal{E})=\\
 &=  \sum_{p,q}\sum_{q'}\left|\mathrm{Tr}\left(\op{U}_\spa{C}^\dagger \op{W}_{q'}^\dagger \op{A}_{p_{q'}^*}\sqrt{a_{p,q}}\op{A}_p\op{W}_q\op{U}_\spa{C}\frac{1}{2^k}\right)\right|^2\\
&=\sum_{q}a_{p_q^*,q}\ .
\end{split}
\end{equation}
We used that $\op{U}_\spa{C}\op{U}_\spa{C}^\dagger = \op{P}_\spa{C}$, $\op{W}_{q'}^\dagger \op{W}_q = \delta_{q,q'} \op{P}_\spa{C}$ and the logical Pauli operators are mutually orthogonal in terms of Hilbert-Schmidt scalar product, $\mathrm{Tr}\big(\op{P}_\spa{C} \op{A}_p \op{A}_{p'} \big) = \delta_{p,p'}\dim(\spa{C}) = \delta_{p,p'} 2^k$. By this formula, the entanglement fidelity in the case of Pauli channels is thus the sum of the $a_{p_q,q}$ probabilities of the errors selected for correction in each syndrome subspace. It is evident, that we get the maximal entanglement fidelity, if we design the error correction operation to correct the error with the greatest probability in each syndrome subspace. Strictly speaking, we have proved only that among the $(2^{2k})^{2^{n-k}}$ different QEC operations which can be given in the Kraus form 
\begin{equation}
\label{eq:all.pauli.correction}
\{\op{U}_\spa{C}^\dagger \op{W}_q^\dagger \op{A}_{p_q}\}_{q=0}^{2^{n-k}-1}\ ,
\end{equation}
there does not exist any better than the one with $p_q=p_q^*$. However, there is not any kind of QEC operation which is better. A complete proof can be found in \cite{fletcher_phd}.

\subsection{Comparison of the standard and optimal QEC}
\begin{figure}
\centering
\includegraphics{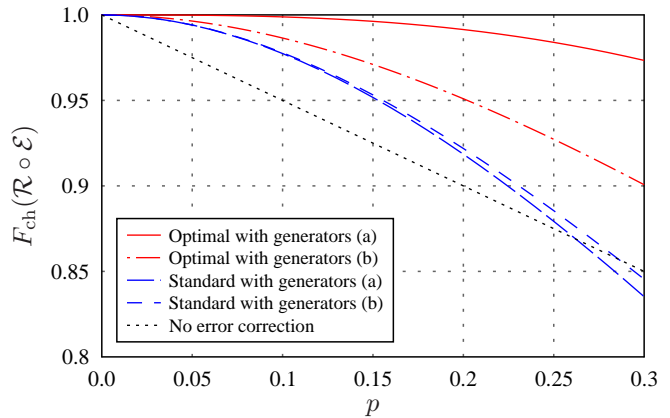}
\caption{(Color online) The channel fidelity for the phase-damping channel using the code in Table \ref{tab:5.qubit.generators}\subref{tab:5.qubit.generators-div} and (i) the optimal error correction operation: solid curve, third order in $p$; (ii) the standard QEC operation: long-dashed line, second order in $p$; using the code in Table \ref{tab:5.qubit.generators}\subref{tab:5.qubit.generators-laf} and (iii) the optimal error correction operation: dotted-dashed line, second order in $p$; (iv) the standard QEC operation: short-dashed line, second order in $p$; finally the case of no error correction (dotted line) is also included for comparison. The optimal correction is better on the full domain of the noise parameter $0 \leq p \leq 1$ not only than the standard curves but also than the no error correction curve.\label{fig:phase.damping.3rd}}
\end{figure}

We shortly mention the cases in which the correction operation based on optimization could give significantly better result than the standard strategy. In these cases one should study carefully which are the most probable errors. When 
\begin{itemize}
\item[(i)] the noise does not act independently on each qubit, i.e. the interaction operator is not the tensor product of single-qubit operators; 
\item[(ii)] the noise is not weak; 
\item[(iii)] even though the channel is tensor product and the noise is weak, it can still occur that a single qubit error has smaller probability than a several qubit error.
\end{itemize} 
A good example on this third case is the phase damping channel, in which $X$ and $Y$ type errors do not occur at all. We use the representation of this channel given by the Kraus operators
\begin{equation}
\label{eq:phase.damping.op.elements}
\op{E}_0=\left(1-\frac{p}{2}\right)I\ ,\ \op{E}_1=\left(\frac{p}{2}\right)Z\ .
\end{equation}
Therefore if we use against this noise the five qubit code \cite{divincenzo} which can be defined by the generators from Table \ref{tab:5.qubit.generators}\subref{tab:5.qubit.generators-div} (see \cite{calderbank.shor_stabilizer}), and according to the standard strategy we use fifteen syndrome subspaces to correct all the fifteen different single-qubit errors then we get significantly worse channel fidelity than in the optimal case, when we use the $2 \times 5$ syndrome subspaces originally designed to correct the single-qubit $X$ and $Y$ errors to correct the two-qubit $Z$ errors. Actually, all the two-qubit $Z$ errors can be corrected in this way by this five qubit code. The result can be seen in Figure \ref{fig:phase.damping.3rd} where the solid line shows the channel fidelity computed by the optimized QEC operation $\spa{R}^*$, this curve is cubic in the noise parameter $\gamma$ in contrast to the long-dashed curve which is only quadratic, showing the channel fidelity computed by the standard QEC operation.

\begin{table}
\subtable[]
{
    \begin{tabular}{c|c}
    \hline\hline
    Name & Operator\\
    \hline
    $g_1$ & $ZXXZI$\\
    $g_2$ & $IZXXZ$\\
    $g_3$ & $ZIZXX$\\
    $g_4$ & $XZIZX$\\
    $\bar{Z}$ & $ZZZZZ$\\
    $\bar{X}$ & $XXXXX$\\
    \hline\hline
    \end{tabular}
    \label{tab:5.qubit.generators-div}
}
\subtable[]
{   
    \begin{tabular}{c|c}
    \hline\hline
    Name & Operator\\
    \hline
    $g_1$ & $XIXZX$\\
    $g_2$ & $ZXZIX$\\
    $g_3$ & $XYZYI$\\
    $g_4$ & $XXIXZ$\\
    $\bar{Z}$ & $IXIZX$\\
    $\bar{X}$ & $XXYYX$\\
    \hline\hline
    \end{tabular}
    \label{tab:5.qubit.generators-laf}
}
\caption{Generators and logical Pauli operators for the five qubit codes.\label{tab:5.qubit.generators}}
\end{table}

Moreover, there is difference between the standard and optimal QEC from another point of view. We can also use the five qubit code developed by Laflamme et al. \cite{laflamme} which can be defined by the generators of Table \ref{tab:5.qubit.generators}\subref{tab:5.qubit.generators-laf} to correct the errors of the phase damping channel. The two codes defined in Table \ref{tab:5.qubit.generators}\subref{tab:5.qubit.generators-div} and \subref{tab:5.qubit.generators-laf} are essentially the same in the sense of the standard strategy, i.e. both are [[5,1,3]] codes which correct perfectly all one-qubit errors. However, the optimal QEC operations based on these codes are very different. As can be seen from the dotted-dashed line in Figure \ref{fig:phase.damping.3rd}, using the code from Table \ref{tab:5.qubit.generators}\subref{tab:5.qubit.generators-laf} the correction of all two-qubit errors are not possible, therefore the curve is only quadratic. In fact the difference of the two codes also can be seen even from the curves of standard QEC (short- and long-dashed lines), because the two codes correct different subset of the higher order (three- and more-qubit) errors. 

Contrary to the above mentioned three cases, the standard strategy is acceptable in most cases, when the channel is not very asymmetric, i.e. the probability of each single-qubit error is about the same. In this case it is easy to see that the most likely errors will be the single-qubit errors, and the traditional QEC correction operation is just designed to correct these, thus it is clearly optimal. A good example on this is the depolarizing channel, for which the optimal entanglement fidelity just equals with the value obtained from the conventional QEC error correction. This result---considered surprising by the authors of \cite{reimpell}---is obvious in the light of the above.

\section{Robustness of error corrections on Pauli channels}
\label{sec:mixing.pauli.channels}

Now we draw conclusions about the robustness from the results of the previous section. For simplicity we assume that $\cha{E}$, $\bar{\cha{E}}$ and so $\cha{E}'_\gamma$ are Pauli channels, so we can directly apply the analytical results of the previous section. Therefore the operator elements from eq. \eqref{eq:mixed.channel.op.elements} can be written equivalently as
\begin{equation}
\label{eq:mixed.pauli.channel.operator.elements}
\left\{\sqrt{(1-\gamma)a_{pq}+\gamma\bar{a}_{pq}}\cdot \op{A}_p\op{W}_q\right\}_{p,q}\ ,
\end{equation}
where $a_{p,q}$ and $\bar{a}_{p,q}$ are the numerical coefficients of the reduced operator elements of the channels $\cha{E}$ and $\bar{\cha{E}}$, respectively, and $a'_{p,q} \coloneqq (1-\gamma)a_{p,q} + \gamma\bar{a}_{p,q}$ are the  numerical coefficients of the reduced operator elements of the channel $\cha{E}'_\gamma$. As can be seen from the formula \eqref{eq:robustness.definition}, to obtain information about the robustness we need the optimal correction for the mixed channel $\cha{E}'_\gamma$. Thus we apply the expression \eqref{eq:entanglement.fidelity.analytical.form}. For this we need to determine the most probable one among the errors which move the code subspace $\spa{C}$ into a given syndrome subspace $\spa{S}_q$, i.e. the greatest $a_{p,q}$ for a given $q$. 

To see what exactly happens in each syndrome subspace with the optimal correction operation during mixing, let us first take a simple case, $\bar{a}_{p,q} = \delta_{p,\bar{p}} \delta_{q,\bar{q}}$ where $\bar{p}$ and $\bar{q}$ are fixed. In this case only in the syndrome subspace $\spa{S}_{\bar{q}}$ will there be any change in the ratios of the $a_{p,q}$ probabilities, so there is no need to examine any other syndrome subspace, as in those the most probable error remains unchanged during the mixing. With the change of the $\gamma$ parameter in the mixed channel only the probability $a'_{\bar{p},\bar{q}}$ grows, the other probabilities are decreasing. There are three cases:
\begin{itemize}
\item[(i)] $a_{\bar{p},\bar{q}}$ is smaller than some other $a_{p^*_{\bar{q}},\bar{q}}$, i.e. in the channel $\cha{E}_0$ it is not the $\op{A}_{\bar{p}} \op{W}_{\bar{q}}$ error operator belonging to the channel $\bar{\cha{E}}$ which is corrected by the current optimal correction, $\cha{R}_0^*$. In this case, however the probability of the error $\op{A}_{\bar{p}} \op{W}_{\bar{q}}$ is increasing as $(1-\gamma) a_{\bar{p},\bar{q}} + \gamma$ while we mix the channels, but the optimal correction remains \emph{unchanged} until the decreasing probability $(1-\gamma) a_{p^*_{\bar{q}},\bar{q}}$ remains greater.
\item[(ii)] $a_{\bar{p},\bar{q}}$ is uniquely the greatest error probability in the syndrome subspace $\spa{S}_{\bar{q}}$. In this case it will increase further during the mixing while the other error probabilities will decrease, so $\op{A}_{\bar{p}} \op{W}_{\bar{q}}$ will be the most probable error and the optimal correction will be unchanged during the whole mixing.
\item[(iii)] Lastly, it can occur that $a_{\bar{p},\bar{q}}$ is one of the greatest error probabilities of $\cha{E}_0$ in the syndrome subspace $\spa{S}_{\bar{q}}$, i.e. there is more than one error operator with the same probability. (In other words, at least two different indices $p_{\bar{q}}^{*} \neq p_{\bar{q}}^{**} \neq \dots$ exist, for which $a_{p_{\bar{q}}^{*},\bar{q}}=a_{p_{\bar{q}}^{**},\bar{q}}=\dots$, and $p_{\bar{q}}^{*}=\bar{p}$.) In this case the optimal correction of the channel $\cha{E}_0$ is not unique. We can choose to correct any of the most probable errors and we get the same fidelity by these completely different correction operations. But when the channel is altered by $\bar{\cha{E}}$, for an arbitrary small value of $\gamma$ the optimal correction operation will uniquely be the one which corrects the error $\op{A}_{\bar{p}} \op{W}_{\bar{q}}$.
\end{itemize}

We can summarize the three cases in the following statement. Supposing that $\cha{R}_0^*$---which is designed to correct the error $\op{A}_{p^*_{\bar{q}}} \op{W}_{\bar{q}}$---is an optimal correction operation for the channel $\cha{E}_0$. Then it is also optimal correction for the mixed channel $\cha{E}'_\gamma$ if $\bar{p} \neq p^*_{\bar{q}}$ and the inequality
\begin{equation}
\label{eq:extremal.pauli.breakpoint.equation}
(1-\gamma)a_{p^*_{\bar{q},\bar{q}}} \geq (1-\gamma)a_{\bar{p},\bar{q}}+\gamma
\end{equation}
holds, or if $\bar{p} = p^*_{\bar{q}}$ and $\gamma$ is arbitrary.

To examine the robustness domain of an error correction operation $\cha{R}_0$ we need to study the fidelities in eq. \eqref{eq:robustness.definition}. First we assume that $\cha{R}_0 = \cha{R}_0^*$ is the optimal error correction of the original channel $\cha{E}_0$. It does not depend on $\gamma$ under the conditions stated in the previous paragraph, therefore $\Delta F_\mathrm{ch} = 0$ as can be seen from eq. \eqref{eq:robustness.definition}. Thus, the boundary of the zero-robustness domain is that $\cha{E}'_{\gamma_0}$ channel, for which $\gamma_0$ is the greatest possible solution of inequality \eqref{eq:extremal.pauli.breakpoint.equation},
\[
\gamma_0 = \frac{ a_{p^*_{\bar{q}},\bar{q}} - a_{\bar{p},\bar{q}} }{ a_{p^*_{\bar{q}},\bar{q}} - a_{\bar{p},\bar{q}} + 1 } \ ,
\]
or $\gamma_0 = 1$ if $\bar{p} = p^*_{\bar{q}}$. For the channel $\cha{E}'_{\gamma_0}$ even the correction operation $\cha{R}_0^*$ is optimal but not uniquely. Furthermore, exceeding this point the optimal correction operation changes abruptly and a completely different error correction will become optimal. Before this $\gamma_0$ boundary point the optimal fidelity goes as
\begin{equation*}
F_\mathrm{ch}(\cha{R}_\gamma^*\circ\cha{E}'_\gamma)=
\sum_{q}(1-\gamma)a_{p_q^*,q}=
(1-\gamma)F_\mathrm{ch}(\cha{R}_0^*\circ\cha{E}_0)\ ,
\end{equation*}
which is the same as the linear curve in \eqref{eq:fidelity.linear.term}, so the $\Delta F_\mathrm{ch}(\gamma)$ defined by \eqref{eq:robustness.definition} is identically zero. (We used that $F_\mathrm{ch}(\cha{R}_0^*\circ\bar{\cha{E}}) = 0$.) But when $a'_{\bar{p},\bar{q}}$ exceeds $a'_{p^*_{\bar{q}},\bar{q}}$, i.e. instead of the correction of the error $\op{A}_{p_{\bar{q}}^*} \op{W}_{\bar{q}}$, the correction of $\op{A}_{\bar{p}} \op{W}_{\bar{q}}$ will be the optimal solution, and the optimal fidelity goes as
\begin{equation}
\label{eq:ent.fidelity.extremal.pauli.after.breakpoint}
F_\mathrm{ch}(\cha{R}_\gamma^*\circ\cha{E}'_\gamma)=\sum_{q\neq\bar{q}}(1-\gamma)a_{p_q^*,q}+(1-\gamma)a_{\bar{p},\bar{q}}+\gamma
\ .     
\end{equation}
As can be seen comparing this equation and eq. \eqref{eq:fidelity.linear.term}, the boundary of the $\delta$-robustness domain for a $\delta > 0$ is that $\cha{E}'_{\gamma_\delta}$ channel, for which 
\[
     \gamma_\delta 
  =  \gamma_0 + \frac{\delta}{a_{p^*_{\bar{q}},\bar{q}} - a_{\bar{p},\bar{q}} + 1}
\ .
\]
In contrast, for every other $\cha{R}_0$ correction operation of type \eqref{eq:all.pauli.correction} which is not optimal for the $\cha{E}_0$ channel, $\gamma_\delta\sim\delta$ without constant term. 

In general when we mix the starting Pauli channel $\cha{E}_0$ with an arbitrary other Pauli channel $\bar{\cha{E}}$ which has more operator elements belonging to different $p,q$ index pairs, then in a given $\spa{S}_q$ syndrome subspace all error probabilities can vary differently, and this happens in more than one syndrome subspace at the same time. In this case the originally optimal error correction operation which is designed to correct the errors $\op{A}_{p^*_q} \op{W}_q$ in each syndrome subspace $\cha{S}_q$ remains optimal also for the mixed channel $\cha{E}'_\gamma$ as long as the inequalities
\begin{equation}
\label{eq:general.pauli.breakpoint.equation}
(1-\gamma)a_{p^*_q,q} + \gamma\bar{a}_{p^*_q,q} \geq (1-\gamma)a_{p,q} + \gamma\bar{a}_{p,q}
\end{equation}
hold for all $p$ and $q$. The greatest value of $\gamma$ for which they are all true defines a $\gamma_0$ for which $\cha{E}'_{\gamma_0}$ is the boundary of the zero-robustness domain of $\cha{R}_0^*$. Going through this point the optimal error correction operation changes abruptly and the fidelity goes linearly with $\gamma$. Therefore, the boundary of the $\delta$-robustness domain for a $\delta > 0$ is that $\cha{E}'_{\gamma_\delta}$ channel, for which $\gamma_\delta = \gamma_0 + \mathrm{const}\cdot\delta$. In contrast, for every other $\cha{R}_0$ correction operation of type \eqref{eq:all.pauli.correction} which is not optimal for the $\cha{E}_0$ channel, $\gamma_\delta\sim\delta$. It is one of our main conclusions that the optimal error correction does not only give better entanglement fidelity but it remains robust in general, except for the case when the greatest common solution of eqs. \eqref{eq:general.pauli.breakpoint.equation} $\gamma_0$ is zero.

Using the following geometric picture we can easily determine and visualize the boundaries of the zero-robustness regions, in which the optimal $\cha{R}^*$ correction operation is the same. Since all the Pauli channels which has the same effect on the code subspace can be given by the numbers $0 \leq a_{p,q} \leq 1$, and because of the trace preserving condition $\sum_{p,q} a_{p,q} = 1$, all of these channels can be represented by a point of a $2^{n+k}-1$ dimensional simplex which has $2^{n+k}$ extremal points. By definition, we assign the perfectly correctable channels as the extremal point at the origin of the simplex. These are the channels which can be reduced to $\op{A}_0 \op{W}_0$ since $\spa{S}_0 = \spa{C}$ and so $\op{W}_0 = I$ and also $\op{A}_0 =I$. The other extremal points of the simplex correspond to the channel reduced to $\op{A}_p \op{W}_q$, $(p,q) \neq (0,0)$. 

A channel $\cha{E}_0$ has unique optimal correction operation $\cha{R}_0^*$ iff the $p_q^*$ indices are uniquely determined for all $q$ by the strict inequalities $a_{p_q^*,q} > a_{p,q}$ for all $q$ and $p \neq p_q^*$. Mixing $\cha{E}_0$ with another channel $\bar{\cha{E}}$ the optimal error correction operation will remain $\cha{R}_0^*$ until eqs. \eqref{eq:general.pauli.breakpoint.equation} hold, i.e. $a'_{p^*_q,q}$ is maximal among $a'_{p,q}$. The point given by the $a'_{p,q}$ numbers is on the boundary between two (or more) domains if there exist at least one index $q$ and at least two indices $p^* \neq p^{**}$ for which $a'_{p^*,q} = a'_{p^{**},q} \geq a'_{p,q}$ where $p$ can be arbitrary. These points are on the $2^{n+k}-2$ dimensional plane which is determined by the following $2^{n+k}-1$ points: all the extremal points except $\op{A}_{p^*} \op{W}_q$ and $\op{A}_{p^{**}} \op{W}_q$ and the midpoint between these two latter extremal points. 

\section{Discussion of Pauli and non-Pauli cases}
\label{sec:conclusion}

In connection with Pauli channels and error correction operations based on stabilizer codes our main conclusions are the following. In general, \emph{the optimal QEC operation is robust}. It means that the QEC operation which is optimal for a channel $\cha{E}_0$ remains also optimal in completely unchanged form for other channels $\cha{E}'_\gamma$ that can be given by small alteration of $\cha{E}_0$. More precisely, the channels which has the same optimal QEC operation form a zero-robustness domain in the space of all Pauli channels. There are finite number of such domains and they cover the whole space of Pauli channels. If a channel is an interior point of such a domain then the optimal QEC operation is robust against \emph{any} Pauli type alteration of the channel, i.e. mixing the channel with any other Pauli channel $\bar{\cha{E}}$, the boundary point $\cha{E}_{\gamma_0}$ of the zero-robustness domain is in non-zero $\gamma_0$ distance. Moreover, the boundary point of an arbitrary $\delta > 0$  $\delta$-robustness domain is outside the zero-robustness domain therefore $\gamma_\delta > \gamma_0$ and $\lim_{\delta \rightarrow 0} \gamma_\delta > 0$. Summarizing, if we have a more or less reliable channel model to describe the interactions between the system and its environment, then choosing the optimal QEC operation we can protect against the uncertainty of the channel model without any knowledge about it. Only when we are on the boundary between two zero-robustness regions is the case worse. In this case there are more (at least two) completely different optimal QEC operations giving the same maximal channel fidelity. By choosing any of them we can find such alteration of the channel---actually that one which could be corrected by the other optimal QEC operation---for which $\gamma_\delta$ starts to increase linearly with $\delta$. In other words there is no such QEC operation which is resistant against any type of channel alteration. If we do not know something about the uncertainty of the channel, then we can not protect against this uncertainty. However, this case is exceptional, most of the channels are interior points of a zero-robustness domain.

In contrast, for an arbitrary channel, in the case of error correction with any other correction operation which has the form \eqref{eq:all.pauli.correction} except the optimal error correction, the $\gamma_\delta$ value of the $\delta$-robustness region starts to increase linearly with $\delta$. This is also true for any other type of QEC operation, having a growth with higher powers of $\delta$ at most in special cases. Based on this we finally can conclude that \emph{for Pauli channels the optimal $\cha{R}^*$ correction operation is better not only in entanglement fidelity but also in robustness than any other correction operation}, assuming the channel is not on the boundary surface of a zero-robustness domain. Therefore we should be careful when the standard QEC differs from the optimal. The latter one takes slight assumptions about the channel, but it does not guarantee the robustness at all. Surprisingly the channel specific optimal solution can give better guarantee for the robustness.

These results are valid for Pauli channels and stabilizer codes, which is a toy model for the general case, but based on some numerical results we believe that the main conclusions are similar also in the case of arbitrary channels. To demonstrate the general situation we study in detail a few examples. For this purpose we choose two non-Pauli channel models, the amplitude-damping channel, and the less familiar ``pure states rotation'' channel defined by Fletcher et al. in \cite{fletcher_near.opt}. The amplitude damping channel is probably the most well known non-Pauli channel. The single qubit operator elements of this channel are the following:
\[
\op{E}_{0}=
\begin{pmatrix}
1 & 0 \\
0 & \sqrt{1-p}\\ 
\end{pmatrix}\ ,\ 
\op{E}_{1}=
\begin{pmatrix}
0 & \sqrt{p} \\
0 & 0 \\ 
\end{pmatrix}\ .
\]
This channel model is physically well motivated. It describes the decay of a state $\ket{1}$ into the state $\ket{0}$. The states $\ket{1}$ and $\ket{0}$ can be excited and ground states of an atom, or a one-photon state and zero-photon state of a single optical mode, so $p$ is the probability that the excited state has decayed to the ground state, or a photon has scattered out from the optical mode.

The pure states rotation channel is in contrast a physically much less motivated channel model. It serves for us only as an example for a non-Pauli channel besides the amplitude-damping, broadening the scope of our results. The single qubit operator elements are the following:
\[
\op{E}_{0}=\alpha
\begin{pmatrix}
\frac{\cos\left(\frac{\theta-\phi}{2}\right)}{\cos\left(\frac{\theta}{2}\right)} & 0 \\
0 & \frac{\sin\left(\frac{\theta-\phi}{2}\right)}{\sin\left(\frac{\theta}{2}\right)}\\ 
\end{pmatrix}\ ,
\]
and
\[
\op{E}_{\pm}=\beta
\begin{pmatrix}
\cos\left(\frac{\theta-\phi}{2}\right)\sin\left(\frac{\theta}{2}\right) & \pm\cos\left(\frac{\theta-\phi}{2}\right)\cos\left(\frac{\theta}{2}\right) \\
\pm\sin\left(\frac{\theta-\phi}{2}\right)\sin\left(\frac{\theta}{2}\right) & \sin\left(\frac{\theta-\phi}{2}\right)\cos\left(\frac{\theta}{2}\right) \\ 
\end{pmatrix}\ .
\]
Here $\alpha$ and $\beta$ are determined by the trace preserving condition. In this channel the parameter $\phi$ characterizes the strength of the noise. Regarding the value of the other parameter, for the simplicity, in this paper we use only the $\theta=\frac{5\pi}{12}$ value.

\begin{figure*}
\subfigure[]
{   
    \includegraphics{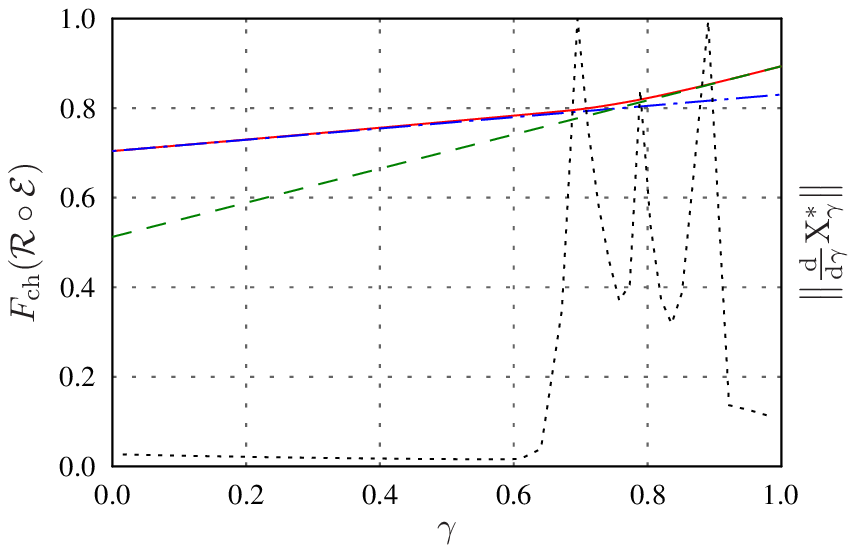}
    \label{fig:pauli.vs.nonpauli-a}
}
\subfigure[]
{   
    \includegraphics{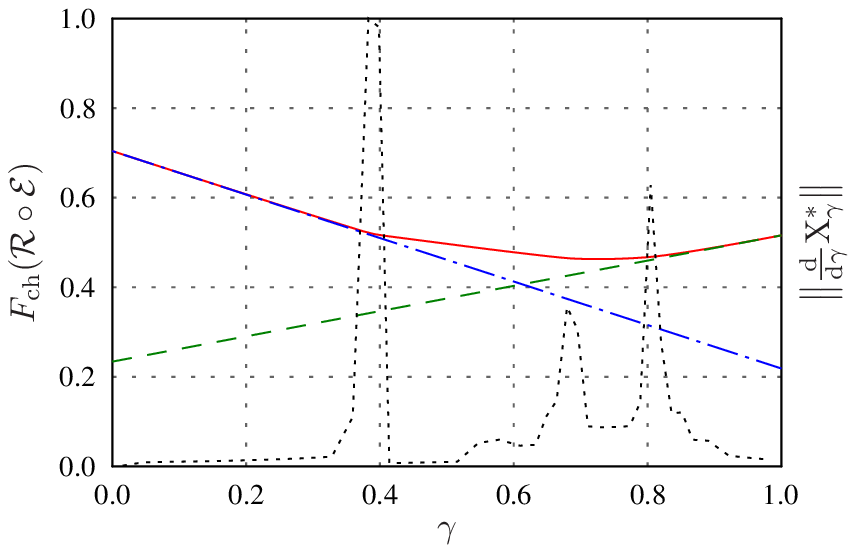}
    \label{fig:pauli.vs.nonpauli-b}
}\\
\subfigure[]
{   
    \includegraphics{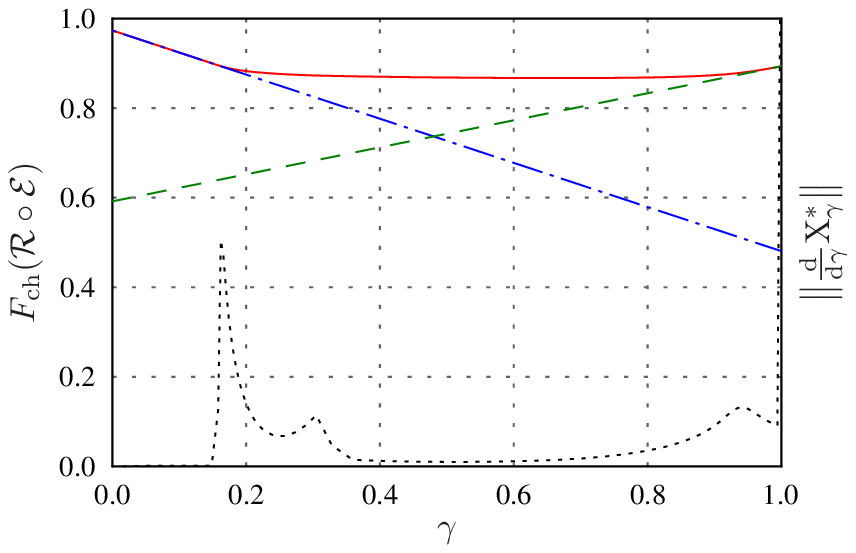}
    \label{fig:pauli.vs.nonpauli-c}
}
\subfigure[]
{   
    \includegraphics{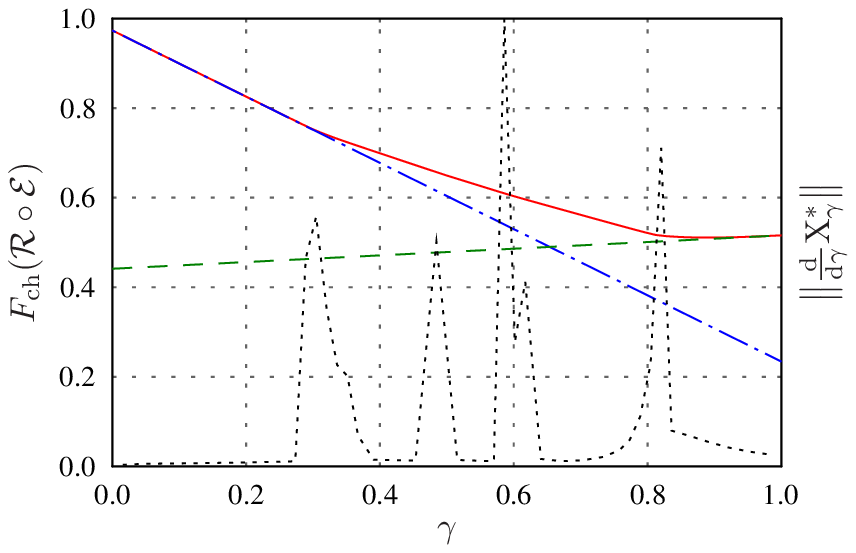}
    \label{fig:pauli.vs.nonpauli-d}
}
\caption{(Color online) The two terms of the $\Delta F$ robustness in function of $\gamma$ for the mixing of Pauli and non-Pauli channels. $\gamma=0$ denotes the initial Pauli channel on each figure. The mixtures are \subref{fig:pauli.vs.nonpauli-a} the depolarizing channel ($p=0.3$) and the amplitude damping channel ($p=0.3$); \subref{fig:pauli.vs.nonpauli-b} the depolarizing channel ($p=0.3$) and the pure states rotation channel ($\theta=\frac{5\pi}{12},\phi=\frac{5\pi}{36}$); \subref{fig:pauli.vs.nonpauli-c} the phase damping channel ($p=0.3$) and the amplitude damping channel ($p=0.3$); \subref{fig:pauli.vs.nonpauli-d} the phase damping channel ($p=0.3$) and the pure states rotation channel ($\theta=\frac{5\pi}{12},\phi=\frac{5\pi}{36}$). The solid line denotes the optimal channel fidelity term, the dashed and dotted-dashed lines denote the linear terms for the two initial channels, and the dotted line denotes the approximate derivative $\big\Vert\dif{\op{X}_\gamma^*}{\gamma}\big\Vert$, which is normalized to its maximum.\label{fig:pauli.vs.nonpauli}}
\end{figure*}

As we mentioned previously, in case of non-Pauli channels we have no analytical method to determine the optimal error correction operation. Therefore hereafter we follow the literature \cite{audenaert,yamamoto,fletcher} and based on the observation that for a fix encoding it is a special convex optimization problem (a semidefinite program), we use numerical methods instead. In our examples, as encoding operation we use the standard five qubit code defined by the generators of Table \ref{tab:5.qubit.generators-div}.

As a fist step, we study the robustness of the optimal QEC operation of a Pauli channel against a non-Pauli altering. We have studied four different cases by mixing two types of Pauli channels with two types of non-Pauli ones. As Pauli channels we use the phase damping channel (see eq. \eqref{eq:phase.damping.op.elements}) and the depolarization channel. We use the representation of the depolarization channel given by the Kraus operators
\[
\label{eq:depolarizing.op.elements}
\op{E}_0=\left(1-\frac{3p}{4}\right)I\ ,\ \op{E}_{k}=\left(\frac{p}{4}\right)\sigma_k\ ,\ k=x,y,z\ .
\]

In the above mentioned four cases we studied systematically the dependence of our results on the noise parameters of the two initial channels that were mixed. We found that the picture is similar in each case. There are no qualitative differences in the results while we modify the noise parameters of the depolarizing, phase damping and amplitude damping channels between zero and $0.7$ and the parameter $\phi$ of the pure states rotation channel between $0$ and $\frac{5\pi}{24}$. In Figure \ref{fig:pauli.vs.nonpauli} we present typical results for all of the four cases. In the figures, the first and the second terms of the expression $\Delta F_\mathrm{ch}(\gamma)$ can be seen, that is the optimal fidelity $F_\mathrm{ch}(\cha{R}^*_\gamma\circ\cha{E}'_{\gamma})$ (solid line) and the linear term $F_\mathrm{ch}(\cha{R}_0\circ\cha{E}'_{\gamma})$ (dotted-dashed line), respectively. It can be seen that the two curves are almost identical at a neighborhood of the initial Pauli channel. (Really, close to the Pauli channel, there is no difference in the two curves in the order of numerical error.) Nevertheless we can not determine exactly the border of the zero-robustness domain, and in principle $\delta_0$ could be zero, but $\delta_\gamma$ starts to increase very slowly with $\gamma$, and by no means linearly. This region could be the generalization of the zero-robustness domain of the pure Pauli case. Increasing the distance from the original Pauli channel we leave this domain and reach another domain where $\delta_\gamma$ clearly has a linear term as a function of $\gamma$. Between these domains there is a transitional region which could be the generalization of the boundary surfaces of the zero-robustness domains. To a first approximation we could summarize our observations based on the analysis of these fidelity curves, that the longer the period which the $F_\mathrm{ch}(\cha{R}^*_\gamma\circ\cha{E}'_{\gamma})$ curve slicks to the linear term, the more we can consider the corresponding correction operation robust against the given alteration. General robustness domains would be those domains, in which the optimal correction operation changes slowly with the alteration of the channel, and the borders between the regions would be those parts, where it changes faster. The necessary condition for the possibility of robust error correction would be that the channel is inside a robustness region. 

To analyze in more  detail the above observations and to clarify the differences and similarities with the pure Pauli case we study the change of the optimal QEC operation with the mixing parameter $\gamma$. In the Pauli case it was completely unchanged in the zero-robustness domain but it was changed abruptly at the boundary of the domain. In the general case, to measure how fast is the change in the optimal QEC operation $\cha{R}_\gamma^*$ we compute the $\big\Vert\dif{\op{X}_\gamma^*}{\gamma}\big\Vert$ derivative of the Choi matrix $\op{X}_\gamma^*$ which is the matrix representative of the superoperator describing the QEC operation $\cha{R}_\gamma^*$. Moreover, $\Vert\op{A}\Vert\coloneqq\sqrt{\mathrm{Tr}(\op{A}^\dagger\op{A})}$ denotes the Hilbert--Schmidt norm. We compute this derivative approximately, it is plotted by dotted line in Figures \ref{fig:pauli.vs.nonpauli}\subref{fig:pauli.vs.nonpauli-a}-\subref{fig:pauli.vs.nonpauli-d}. We observe that $\cha{R}_\gamma^*$ varies very slowly near the Pauli channel, i.e. if $\gamma$ is small. Even more interestingly we can see that there are abrupt changes in $\cha{R}_\gamma^*$, denoted by the peaks of the dotted curves. The better the numerical precision, the higher these peaks are, indicating discontinuous jumps in the $\cha{R}_\gamma^*$ optimal QEC operation. Between these non-continuous changes $\cha{R}_\gamma^*$ is not constant but changes relatively slowly as a smooth function of $\gamma$. This indicates that the optimal QEC operation changes in two different ways. The first case is similar to the domain boundary of the Pauli case, where in a syndrome subspace one of the error operators became more likely than the others. We believe that this similarity can not be accidental, however at this moment we can not see how it could be precisely generalized to the non-Pauli case.

We also have to mention that there are more breakpoints in the optimal fidelity curve which can not be seen because of the scaling of the figure. However, changing the scale and enlarging the vicinity of a point where $\cha{R}_\gamma^*$ is discontinuous, we always observe a small breakpoint in the optimal fidelity curve. It is important to note that between these breakpoints, where $\cha{R}_\gamma^*$ varies analytically, the optimal QEC operation is robust in the sense that small change in $\gamma$ causes a change of only second order in $\Delta F$. But in the non-analytic points this is not true, the optimal $\cha{R}_\gamma^*$ is not uniquely defined and there is no good choice against an arbitrary channel alteration. These points however form a discrete set in the $\gamma \in [0,1]$ interval, and a subset of zero measure in the space of all possible channels. In the light of the above results we can assert that \emph{the optimal QEC operation of a Pauli channel is also robust against non-Pauli channel perturbations}.

\begin{figure}
\centering
\includegraphics{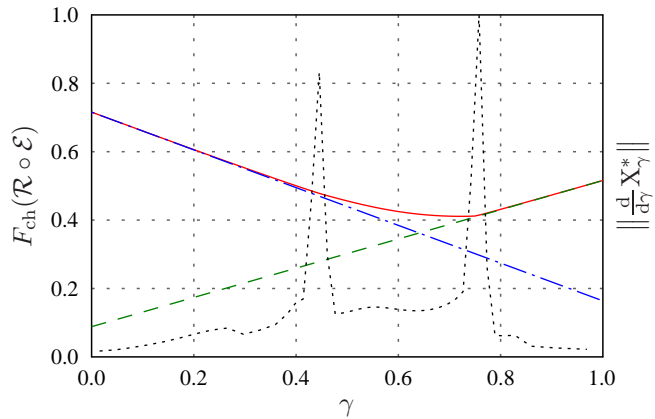}
\caption{(Color online) The two terms of $\Delta F$ in function of $\gamma$ for the mixing of the amplitude damping channel ($p=0.5$) and the pure states rotation channel ($\theta=\frac{5\pi}{12},\phi=\frac{5\pi}{36}$), and the approximate derivative $\big\Vert\dif{\op{X}_\gamma^*}{\gamma}\big\Vert$. Notations are similar to those of Figure \ref{fig:pauli.vs.nonpauli}.\label{fig:amp.vs.prot}}
\end{figure}

\begin{figure}
\centering
\includegraphics{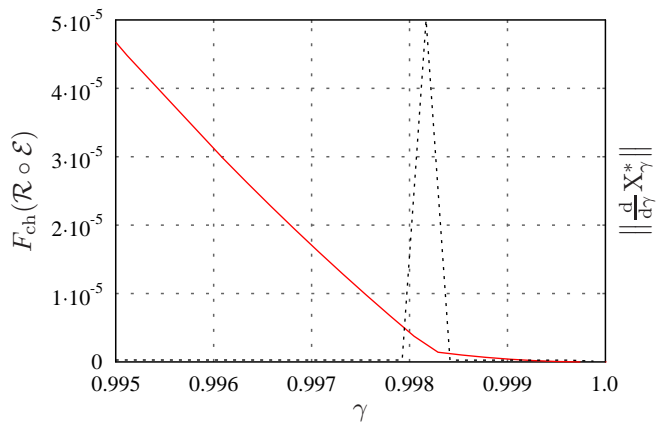}
\caption{(Color online) The enlarged version on Figure \ref{fig:pauli.vs.nonpauli-c} in the small vicinity of $\gamma=1$. The second term of $\Delta F$ for the non-Pauli side was subtracted from the first, thus the red line shows the $\Delta F$ function. The dotted line is again the normalized approximate derivative $\big\Vert\dif{\op{X}_\gamma^*}{\gamma}\big\Vert$. \label{fig:pha03-amp03-break}}
\end{figure}

On the other hand, the curves of Figure \ref{fig:pauli.vs.nonpauli} can also be viewed from the other side, i.e. from the $\gamma = 1$ endpoint. This point describes the initial amplitude damping or pure states rotation channels, respectively. Going backwards from this point in $\gamma$ we can study the robustness of the QEC operation which is optimal for a non-Pauli channel against a Pauli type alteration. As can be seen from \eqref{eq:fidelity.linear.term}, the second term of eq. \eqref{eq:robustness.definition}, $F_\mathrm{ch}(\cha{R}_0\circ\cha{E}'_{\gamma})$ varies linearly in function of $\gamma$ for an arbitrary channel and an arbitrary alteration, i.e. as the distance form the channel $\cha{E}_0$ grows. This is indicated by the dashed line in Figure \ref{fig:pauli.vs.nonpauli}. In contrast to the case of Pauli channels we can see from the dotted lines that the derivative of the Choi matrix is not zero. This means that the  $\cha{R}_\gamma^*$ optimal QEC operation changes by arbitrary small alteration of the channel. As a consequence $\delta_0=0$, i.e. there is no finite zero-robustness domain around a non-Pauli channel. However, from the fidelity curves we can see that the optimal term $F_\mathrm{ch}(\cha{R}_\gamma^*\circ\cha{E}'_{\gamma})$ slicks to the linear term $F_\mathrm{ch}(\cha{R}_0\circ\cha{E}'_{\gamma})$ and $\gamma_\delta$ starts to increase only with higher powers of $\delta$ and not linearly. Therefore \emph{the optimal QEC operation proved to be more robust than any other QEC operation}, for which the $\gamma_\delta$ value of the $\delta$-robustness region starts to increase linearly with $\delta$. Only when we are in a point where the optimal QEC operation changes abruptly---at the peaks of the dotted line---is the case worse, very similar to the boundaries of zero-robustness domains in the case of Pauli channels. We can see an interesting example in Figure \ref{fig:pauli.vs.nonpauli-c}. Very close to the amplitude damping channel there is a very narrow peak in the derivative $\big\Vert\dif{\op{X}_\gamma^*}{\gamma}\big\Vert$ indicating an abrupt change in the optimal QEC operation, therefore the optimal and the linear lines split at that point. Enlarging the vicinity of the amplitude damping channel (see Figure \ref{fig:pha03-amp03-break}) we observe that the optimal and the linear terms are almost identical when $0.9982\lessapprox\gamma$ and the difference between these terms grows linearly as $\gamma$ decreases, i.e. the distance from the initial amplitude damping channel is increasing. The picture is very similar to the Pauli case, nevertheless the robustness domain is very small.

To make sure that our statements are general, we also studied a case where two non-Pauli channels are mixed. For this purpose we mix the amplitude damping and the pure states rotation channel. Again, we scan the $p\in [0,0.7]$ and $\phi\in [0,\frac{5\pi}{24}]$ noise parameter intervals of the channels. As a typical example we plot the $p=0.5$ and $\phi=\frac{5\pi}{36}$ case in Figure \ref{fig:amp.vs.prot}. Our observations are essentially the same as in the previously discussed case. Actually, we can also see such non-Pauli--non-Pauli channel mixing cases anywhere in the figures of Pauli--non-Pauli cases, when we take two channels corresponding to nonzero $\gamma$ points and mix them.

The results stated above strongly indicate---however not prove---the following consequences. In contrast to the pure Pauli case, zero-robustness domains, where the optimal recovery operation is not changes at all, does not cover the whole space. Instead, $\cha{R}_\gamma^*$ can change in two different ways. The first is similar to the case of boundaries of Pauli zero-robustness domains. In these special points $\cha{R}_\gamma^*$ is not unique, it changes abruptly and the fidelity is non-analytic function of the mixing parameter. Between these points $\cha{R}_\gamma^*$ changes analytically with the mixing parameter like the fidelity, and therefore the optimal QEC operation can said to be robust.

A more precise study of the distribution of the non-analytic points in the space of all quantum channels is the task for the future.

\end{document}